\documentclass[aps,pra,epsfigure,twocolumn,longbibliography]{revtex4-1}
\usepackage{dcolumn}
\usepackage{bm}
\usepackage{graphicx}
\usepackage{amsmath}
\usepackage{latexsym}
\usepackage{amsfonts}
\usepackage{amssymb}
\usepackage{array}
\usepackage{epsfig}
\usepackage{txfonts}
\usepackage{color}
\usepackage[colorlinks=true,linkcolor=blue,urlcolor=blue,citecolor=blue]{hyperref}
\usepackage{bm}
\usepackage{setspace}
\usepackage{braket}
\usepackage[utf8]{inputenc}
\usepackage[percent]{overpic}
\usepackage{xcolor}

\newcommand{\beq}{\begin{equation}}
\newcommand{\eeq}{\end{equation}}
\newcommand{\beqa}{\begin{eqnarray}}
\newcommand{\eeqa}{\end{eqnarray}}

\def\beq{\begin{equation}}
\usepackage{color}

\begin{document}
\title{Trigonometric protocols for shortcuts to adiabatic transport of cold atoms in anharmonic traps}
%\date{\today}

\author{Jing Li}

\author{Qi Zhang}

\author{Xi Chen}
\email{xchen@shu.edu.cn}

\affiliation{Department of Physics, Shanghai University, 200444 Shanghai, People's Republic of China}

\begin{abstract}
Shortcuts to adiabaticity have been proposed to speed up the ``slow" adiabatic transport of an atom or a wave packet of atoms. However, the freedom of the inverse engineering
approach with appropriate boundary conditions provides thousands of trap trajectories for different purposes, for example, time and energy minimizations.
In this paper, we propose trigonometric protocols for fast and robust atomic transport, taking into account cubic or quartic anharmonicities. The numerical results have illustrated that such trigonometric protocols, particular cosine ansatz, is more robust and the corresponding final energy excitation is smaller, as compared to sine trajectories implemented in previous experiments.
\end{abstract}
\maketitle

\section{Introduction}

The accurate manipulation of atomic motion is quite demanding in current quantum science and technique
\cite{HanschNature2001,HanschPRL2001,Ketterle2002,Leibfried,ions,Bowler,Walther},
with applications ranging from basic science, metrology to quantum information processing.
Different from the sufficiently ``slow" adiabatic moving,
several approaches, including optimal control, have been put forward to achieve fast non-adiabatic transport \cite{David08,Calarco09,HaichaoAPL10,YouLi}.
The reduced transport time may make the atom manipulation more productive in practice, and also avoid overheating from coils
and fluctuating fields and decoherence effects.

Recently, the concept of ``shortcuts to adiabaticity" (STA) \cite{reviewSTA13} provides an alternative technique for fast transport, and
faithful to the ideal result of adiabatic transport \cite{Masuda10,Erik11,Xi12,Erikcond12,Mikel13,Uli14,CampoPRX,Lu14,Mikel14,David14,Qi2015,Qi,AdolfNC}.
In particular, the invariant-based inverse engineering, combining perturbation theory and optimal control, is considered as a versatile toolbox
for designing the optimal transport protocols, according to different physical criteria or operational constraints \cite{Xi12,Erikcond12}.
In other word, among the family of shortcuts satisfying the initial and final conditions, the specific
path can be chosen by optimizing the operation time or transient
excitation energy, with a restriction of the allowed transient
frequencies. Furthermore, the fast transport can be further optimized with respect to spring-constant (color) noise, position fluctuation \cite{Lu14}, and spring-constant error \cite{David14}.

On many cold atom or ion experiments, shortcuts to adiabatic transport are mostly designed
for perfectly harmonic traps but most confining traps, \textit{i.e.}, magnetic quadrupole potential \cite{HaichaoAPL10},
gravitomagnetric potential \cite{Ricci}, electrostatic potential \cite{AlonsoNJP} and optical dipole traps \cite{OptTrap}, are of course anharmonic.
Actually, the perturbing effects of anharmonicities are of paramount importance in actual trap \cite{Erik3DPRA},
which implies the unwanted final excitation, or even atom loss.
This sets the physical limits to the possible speed-up, due to the intermediate energy excitation \cite{Chenenergy10}.
In Ref. \cite{LuPRA2014}, the optimal ``bang-singular-bang" control is designed to achieve fast transitionless expansion
of cold neutral atoms or ions in Gaussian anharmonic trap, with minimizing the time-averaged perturbative energy.
In fact, the effects of anharmonicity is also one of significant problems on protocol designing in ion transport \cite{Mikel13},
in which the optimal strategy is strongly required to minimize excitation in presence of anharmonicity.
Up to now, several works have been devoted to dealing with the transport in anharmonic traps and overcoming the difficulty. (i) The trap trajectory of atomic transport in general power-law traps including
cubic or quartic anharmonicities have been calculated from the classical Newton equation, and the quantum case for a wave packet has been checked later \cite{Qi2015}. (ii) The counter-diabatic driving, suggesting
the compensating force, proposed for nonharmonic traps \cite{CampoPRX}, which has been currently implemented for ion transport \cite{AdolfNC}. However the trap frequency and size of atom cloud might be modified when the anharmonicity is present \cite{Qi}. (iii) The combination of inverse engineering and optimal control theory is proposed, but the anharmonic potential is always considered as perturbation \cite{Qi}.

In this article, we put forward the trigonometric protocols for shortcut to adiabatic transport in anharmonic traps, including the cubic or quartic anharmonicities.
Particularly, we try a simple but efficient cosine ansatz with additional boundary condition to eliminate the anharmonic corrections, and finally achieve fast and robust transport of atoms with null final excitation energy. Such excellent cosine protocol has quite remarkable behavior in cancelling the spring constant error for two-ion transport \cite{Lu2015}. It is also
similar to but different from sine protocol implemented in the experiment of atom transport,
in which the high efficiency \cite{YouLi}, above $97 \%$, has been reported. Our numerical simulations have demonstrated that
our designed shortcuts with cosine protocols is more stable with respect to anharmonic effects, and the corresponding final excitation energy
is smaller, as compared to the sine protocols. All results presented here are oriented to the current experiment of transport neutral atoms \cite{David08,HaichaoAPL10,YouLi}, but can be applicable
to the ion transport \cite{Bowler,Walther,AdolfNC}.

%The paper is organized as follows. In Sec.\ref{oneb}, we introduce the inverse-engineering method to transport of non-harmonic systems.

\section{Transport with anharmonic traps}
\label{oneb}
\subsection{cubic anharmonicity}

Consider the transport of an atom of mass $m$, which is confined in a nonharmonic trap. First of all, the cubic anharmonicity is considered, which results from an expansion around the minimum of the real transport potential \cite{David08,AlonsoNJP,Qi2015,Qi}. The whole potential is written as
\begin{equation}
\label{anah}
V(x,t)  =  \frac{1}{2}m\omega_0^2\left(x-x_0(t)\right)^2 +  \frac{1}{3}m\frac{\omega_0^2}{\xi}\left(x-x_0(t)\right)^3,
\end{equation}
where $x_0(t)$ represents the trajectory of the bottom of the trap to be determined.
$\xi$ quantifies the strength of the cubic anharmonicity.
From the expression, we could see this kind of trap has a finite depth and asymmetry.
According to the Newton's law, the motion of particle obeys the following differential equation,
\begin{equation}
\ddot x + \omega_0^2 \left(x-x_0(t)\right) + \frac{\omega_0^2}{\xi}\left(x-x_0(t)\right)^2=0,
\label{eqcubic}
\end{equation}
from which we finally obtain
\begin{equation}
x_0(t) = x(t) +\frac{\xi}{2} \left( 1-\sqrt{1-\frac{4\ddot x}{\xi\omega_0^2}} \right).
\label{eqnl1}
\end{equation}
In this case, an exact strategy for $x(t)$ can be worked out by choosing the appropriate trap trajectory $x_0(t)$ with right boundary conditions based on the inverse engineering approach.
To minimize the effect of anharmonicity, we rewrite and solve perturbatively Eq.~(\ref{eqcubic}), see Ref. \cite{Qi2015},
\beq
\ddot{\tilde{x}}_2 + u^2 (\tilde x_2-\tilde x_0)=-\frac{u^2 d}{\xi} (\tilde x_2-\tilde x_0)^2 \simeq  -\frac{d}{\xi} \frac{(\ddot{\tilde x}_1)^2}{u^2},
\eeq
where $u=\omega_0 t_f$, $\tilde{x}_0=x_0(t)/d$, $s=t/t_f$, and $\tilde{x}_1(t)$ satisfies the Newton equation of perfect harmonic trap
\beq
\ddot{\tilde{x}}_1+u^2[\tilde{x}_1-\tilde{x}_0(s)]=0.
\label{HO}
\eeq
The perturbative solution to the first order is $\tilde x_2(s) = \tilde x_1(s) + (d/\xi)f_1(s)$ with
\begin{equation}
f_1(s)=-\frac{1}{u^3} \int_0^s \ddot{\tilde{x}}^2_1(s') \sin [u(s-s')] ds'.
\label{f1}
\end{equation}
In order to obtain the shortcut to adiabatic transport of atom, one has to nullify the final (dimensional) residual energy at $t=t_f$ \cite{David14,Qi2015,Qi},
\begin{equation}
\frac{\Delta E}{\hbar \omega_0} = \frac{m \omega_0 d^2}{\hbar}
\left[\frac{\dot{\tilde{x}}^2}{2u^2} +\frac{(\tilde x-\tilde x_0)^2}{2} + \frac{d}{3\xi}(\tilde x-\tilde x_0)^3 \right],
\label{energy}
\end{equation}
which gives the boundary conditions at initial and final times, that is,
$\tilde{x}_1(0)=0$, $\dot{\tilde{x}}_1(0)=\dot{\tilde{x}}_1(1)=0$, $\ddot{\tilde{x}}_1(0)=\ddot{\tilde{x}}_1(1)=0$, and $\tilde{x}_1(1)=1$.
These are the same as those imposed from the commutator relation between the dynamical invariant and Hamiltonian at the time edges, when the inverse engeering approach based on
Lewis–Riesenfeld invariant is applied, see Refs. \cite{Erik11,Xi12}. In addition, one more condition, $f_1(s)=0$, could be included to make the
final excitation energy zero, guaranteeing the robust with respect the cubic anharmonicity.

\begin{figure}[]
\centering
\scalebox{0.85}[0.85]{\includegraphics{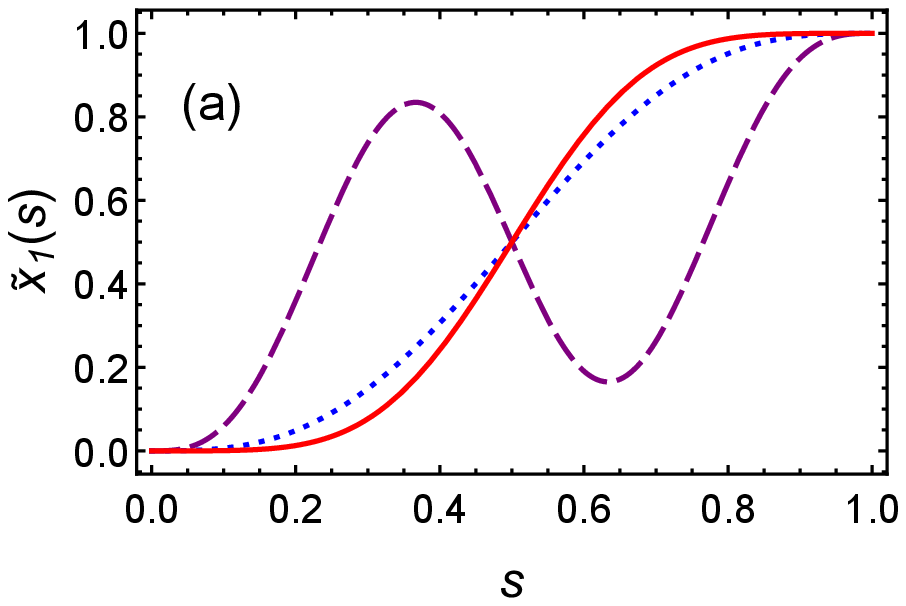}} \\
\scalebox{0.85}[0.85]{\includegraphics{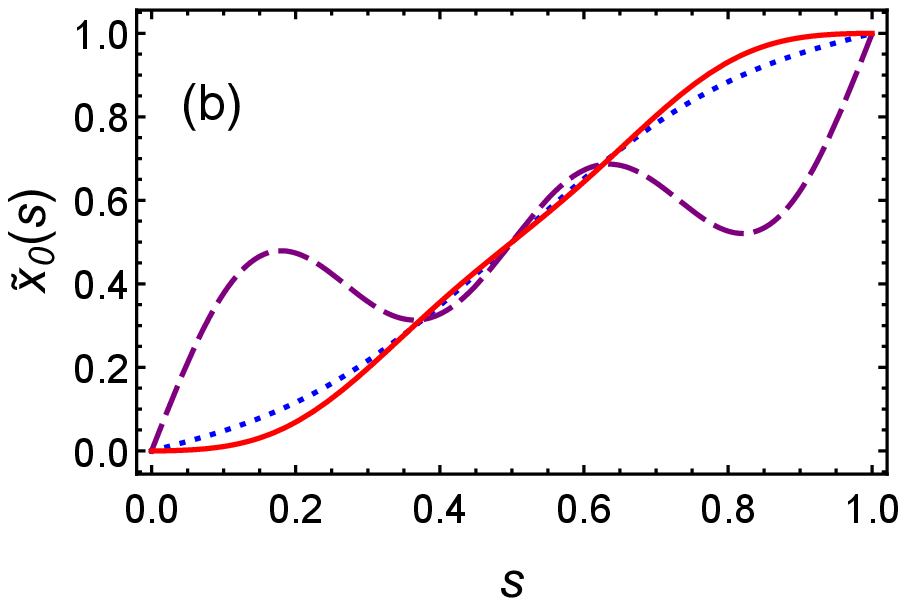}}
\caption{\label{fig1} (Color online) (a) The functions of different trigonometric protocols and (b) corresponding trajectories of trap center, where the cosine ansatz (\ref{xc a1 cos}) (solid red line), sine ansatz (\ref{xc sin}) (dotted blue line) and sine ansatz (\ref{xc a1 sin}) with one more parameter nullifying $f_1(s)$ (dashed purple line). Parameters $u=3\pi$.}
\end{figure}

To this end, we try the trigonometric protocols, particularly, the cosine ansantz, for designing the shortcuts with minimizing the anharmonic effects.
By assuming the cosine ansantz, $\tilde{x}_1(t)=a_0+\sum^3_{j=1}a_j\cos\left[(2j-1)\pi s\right]$,
we solve the trajectory of the center of mass satisfying all six boundary conditions mentioned above, and additional condition $f_1(s)=0$.
Thus one free parameter $a_1$ is added for the mass center, $x_1(s)$, which is
\beqa
\label{xc a1 cos}
\tilde{x}_1(s) = \frac{1}{2} + a_1 \cos(\pi s) + a_2 \cos(3 \pi s) + a_3 \cos(5 \pi s),
\eeqa
with the numbers $a_1=-0.579$, $a_2 = 0.08725$ and $a_3= -0.00825$. In this case,
$\tilde{x}_2(t)=\tilde{x}_1(t)$, the trajectory of trap center $\tilde{x}_0(t)$ can be obtained by solving Eq (\ref{HO}).
In the following discussion, we shall check the stability of designed trajectory $x_0(t)$, (also $\tilde{x}_0(t)$). If one wants to check the final residual energy (\ref{energy}), the solution $x(t)$ can be
directly solved from Eq. (\ref{eqcubic}) with the boundary conditions $x(0)=0$ and $\dot{x}(0)=0$. Remarkably, there are several advantages of the cosine ansatz that we shall emphasize.
On one hand, the trigonometric ansatz with only four free parameters is much simpler than the conventional polynomial ansatz, in which at least seven coefficients should be assumed and solved numerically \cite{Erik11,Xi12}.
On the other hand, the trigonometric ansatz is more efficient to cancel the anharmonic correction, induced from Eq. (\ref{f1}). The residual excitation energy in this case can be reduced by at least one hundred times,
as compared to conventional polynomial ansatz with seven parameters, guaranteeing $f_1 (s)=0$.

For comparison, we also write the simple sine ansatz for $\tilde{x}_1(s)$,
\beq
\label{xc sin}
\tilde{x}_1(s)=s-(1/2\pi)\sin\left(2\pi s \right).
\eeq
This sine protocol is relevant to but slightly different from sine protocol used in the experiment \cite{YouLi}, in which
$\tilde{x}_0(s)=s-(1/2\pi)\sin\left(2\pi s \right)$ is assumed and thus
$\tilde{x}_1(s)=s - (9/10 \pi)\sin\left(2 \pi s\right)$. As a matter of fact, the reason for achieving high fidelity, above $97\%$, is that
the $\tilde {x}_1$ satisfies the boundary conditions $\tilde{x}_1(0)=\ddot{\tilde{x}}_1(0)=\ddot{\tilde{x}}_1(1)=0$ and $\tilde{x}_1(1)=1$. However, the boundary condition
$\dot{\tilde{x}}_1(0)=\dot{\tilde{x}}_1(1)=-0.8 \neq 0$ suggests that it is not exact shortcut protocol.
Moreover, we can also add one more free parameter in sine ansatz for nullifying $f_1$, see Eq. (\ref{f1}), which results in
\beq
\label{xc a1 sin}
\tilde{x}_1(s)=s + a_1 \sin\left(2\pi s \right)+a_2 \sin\left(4\pi s \right),
\eeq
with $a_1 =0.3135$ and $a_2=-0.236348$. Once $\tilde{x}_1(s)$, one can calculate $\tilde{x}_0 (s)$ accordingly from Eq. (\ref{eqcubic}).

\begin{figure}[]
\centering
\scalebox{0.85}[0.85]{\includegraphics{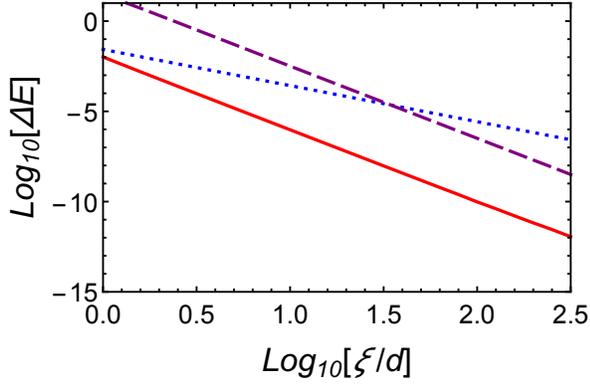}}
\caption{\label{fig2}(Color online) Final residual energy $\Delta E$, defined by Eq. (\ref{energy}), versus the parameter $\xi/d$ for a cubic anharmonicity for different trigonometric protocols, where the cosine ansatz (\ref{xc a1 cos}) (solid red line), sine ansatz (\ref{xc sin}) (dotted blue line) and sine ansatz (\ref{xc a1 sin}) with one more parameter nullifying $f_1(s)$ (dashed purple line).
Parameters: $\omega_0 = 2 \pi \times 1.41 \times10^5$ Hz, $u=3\pi$, $m = 40 \times 1.667 \times 10^{-27}$ Kg ($^{40}$Ca$^+$), $a_0 = [\hbar/(m \omega_0)]^{1/2}$ and $d = 20.2 a_0$.}
\end{figure}

Figure \ref{fig1} (a) shows the function of different trigonometric protocols, obtained from Eqs. (\ref{xc a1 cos})-(\ref{xc a1 sin}), in which
all trigonometric protocols including sine and cosine ansatzes require fewer parameters, as compared to polynomial ansatz, see the examples in Ref. \cite{Erik11}.
For comparison, the final residual energy for cosine ansatz at $t=t_f$ is much smaller than the one for sine ansantz, as shown in Fig. \ref{fig2}, though few parameters are required for
the sine ansantz. This suggests that the consine ansatz for fast transport is more stable with respect with the cubic anharmonicity.
Moreover, the final excitation energy in principle decreases when the effect of anharmonicity becomes weaker, with increasing $\xi$.
Since the sine protocol (\ref{xc a1 sin}) with additional parameter for nullifying $f_1(s)$ is only valid for a perturbation anharmonicity when $\log_{10}{(\xi/d)} >1.51$.
So in the following numerical calculation, we focus on the cosine ansantz (\ref{xc a1 cos}) for fast and robust transport of atoms, comparing with sine one (\ref{xc sin}).

\begin{figure}[]
\centering
\scalebox{0.85}[0.85]{\includegraphics{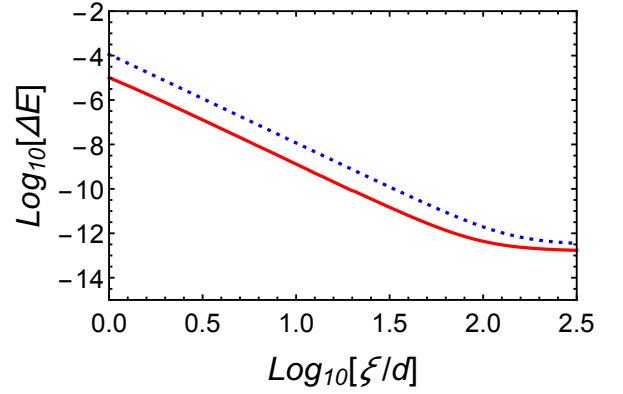}}
\caption{\label{fig3}(Color online) Final residual energy $\Delta E$, defined by Eq. (\ref{energy2}), versus the parameter $\xi/d$ for a quartic anharmonicity for different trigonometric protocols, where the cosine ansatz (\ref{cos2}) (solid red line) and sine ansatz (\ref{xc sin}) (dotted blue line). The parameters are the same as those in Fig. \ref{fig2}.}
\end{figure}

\subsection{Quartic anharmonicity}

Now let us consider another quartic anharmonic trap, which exists frequently in a realistic experiment.
The sum of harmonic potential and quartic anharmonicity form
\beq
V(x,t)=\frac{1}{2}m\omega_0^2\left(x-x_0(t)\right)^2+ \frac{1}{4}m \frac{\omega_0^2}{\xi^2}\left(x-x_0(t)\right)^4.
\eeq
The motion of particle, satisfying the Newton's equation, gives
\beq
\ddot x + \omega_0^2 \left(x-x_0(t)\right) + \frac{\omega_0^2}{\xi^2}\left(x-x_0(t)\right)^3=0.
\label{eqquartic}
\eeq
Repeating the above strategy, we solve perturbatively and obtain
\beq
\ddot{\tilde{x}}_2 + u^2 (\tilde x_2-\tilde x_0)=-\frac{u^2 d^2}{\xi^2} (\tilde x_2-\tilde x_0)^3 \simeq  \left(\frac{d}{\xi}\right)^2 \frac{(\ddot{\tilde x}_1)^3}{u^4},
\eeq
where $u=\omega_0 t_f$, $\tilde{x}_0=x_0(t)/d$, $s=t/t_f$, and $\tilde{x}_1(t)$ satisfies the Newton equation (\ref{HO}).
The perturbative solution to the first order is $ \tilde x_2(s) = \tilde x_1(s) + \left(d/\xi\right)^2 f_2(s)$
with
\beq
f_2(s)=\frac{1}{u^5} \int_0^s (\ddot{\tilde{x}}_1(s'))^3 \sin [u(s-s')] ds'.
\eeq
Again, the residual energy at the final time, $t=t_f$, is calculated as
\beq
\label{energy2}
\frac{\Delta E}{\hbar \omega_0} = \frac{m \omega_0 d^2}{\hbar} \left[ \frac{\dot{\tilde{x}}^2}{2u^2} + \frac{(\tilde x-\tilde x_0)^2}{2} + \frac{d^2}{4\xi^2}(\tilde x-\tilde x_0)^4  \right].
\eeq
Similarly, we assume the cosine ansatz as follows,
\beq
\label{cos2}
\tilde{x}_1(s) = \frac{1}{2} + a_1 \cos(\pi s) + a_2 \cos(3 \pi s) + a_3 \cos(5 \pi s),
\eeq
with the numbers $a_1= -0.513628$, $a_2=-0.0108075$ and $a_3=0.0244358$, satisfying the boundary conditions mentioned above and making $f_2(s)=0$. Once $\tilde{x}_1 (s)$ is fixed, the
trajectory of trap center $x_0(t)$, namely, $\tilde{x}_0 (s)$ can be calculated from the Newton equation (\ref{HO}), and the center of mass, $x(t)$, can be calculated from Eq. (\ref{eqcubic}), as a consequence.
Fig. \ref{fig3} illustrates that the cosine protocol (\ref{cos2}) designed here is better than the simple sine protocol (\ref{xc sin}), where the final residual energy is reduced by
one hundred times for the same parameters in Fig. \ref{fig2}.  Noting that to avoid the singularity in the numerical calculations, we choose $u= 3.00001 \pi$ instead of exact value of $u= 3 \pi$.

\begin{figure}[]
\centering
\scalebox{0.85}[0.85]{\includegraphics{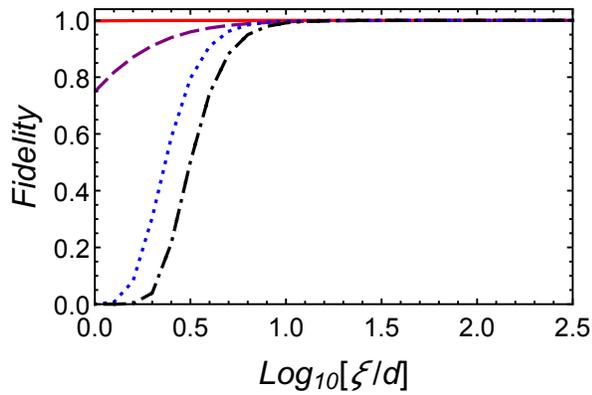}}
\caption{\label{fig4} (Color online) Fidelity versus the anharmonicity parameter $\xi/d$ for cubic and quartic anharmonicities with trigonometric trajectories,
where cosine ansantz (solid red line) and sine ansatz (purple dashed line) for cubic anharmonicity, and cosine ansatz (dotted blue line) and sine ansatz (dash-dotted black line). The parameters are the same as those in Fig. \ref{fig2}.}
\end{figure}

Finally, we shed light into the stability of different the trigonometric protocols, including sine and cosine ansatzes, see Eqs. (\ref{xc a1 cos}), (\ref{xc sin}) and (\ref{cos2}), when the cubic and
quartic anharmonicities are taken into account. To perform the numerical calculation, we start from the eigenstate of harmonic trap at initial time $t=0$, and solve the time-dependent Schr\"{o}dinger equation
with designed trajectory $x_0(t)$ by using spit-operator method. The fidelity is defined as $F= |\langle \psi_0 (t_f)| \tilde{\psi}(t_f) \rangle|^2$, where the desired state $\psi_0 (t_f)$ is the eigenstate of harmonic trap at final time $t=t_f$ with displacement $d$, and $\tilde{\psi}(t_f)$  is the numerical results. Fig. \ref{fig4} demonstrates that the cosine protocols for both cubic and quartic anharmonicities are
better than the sine ones. The improvement of cosine protocol is more pronounced in the case of cubic anharmonicity which is consistent with the results for final residual energies, see Figs. \ref{fig2} and \ref{fig3}, in which
more excitation energy at $t=t_f$ can be reduced.

\section{Conclusion}

To conclude, we have studied the trigonometric protocols including sine and cosine ansantzs for fast and robust atomic transport, taking into account cubic or quartic anharmonicities.
We have found that the simple but efficient cosine protocol is more stable with respect to the parameter of anharmonicity when the final residual energy is nullified.
Such shortcut protocols are applicable in the experiments on fast and robust transport of cold atoms \cite{HaichaoAPL10,YouLi} and ions \cite{AdolfNC}. In addition,
the trigonometric protocols can be also useful for designing the shortcuts to adiabatic compression/expansion in harmonic trap for fast frictionless cooling \cite{PRL104} and
minimizing the excess work in thermally isolated systems \cite{PRE15}.

\section*{acknowledgements}
We thank J. Gonzalo Muga and D. Gu\'{e}ry-Odelin for their fruitful discussions on this topic.
This work was partially supported by the NSFC (11474193),
the Shuguang Program (14SU35), the Specialized Research Fund for the Doctoral Program (2013310811003),
and the Program for Eastern Scholar.

\end{document}